\documentclass[
aps,prapplied,showpacs,superscriptaddress,numerical,amsmath,amssymb,floatfix,reprint,longbibliography]{revtex4-2}
\usepackage[T1]{fontenc}
\usepackage[utf8]{inputenc}
\usepackage[
pdffitwindow=true,
colorlinks=true,
frenchlinks=false,
linkcolor=blue,
anchorcolor=blue,
citecolor=blue,
filecolor=blue,
urlcolor=blue,
bookmarks=true,
bookmarksopen=true,
bookmarksnumbered=true,
bookmarksopenlevel=1,
plainpages=false,
pdfpagelayout=TwoPageLeft,
pdfpagelabels=true,
breaklinks]{hyperref}
\usepackage[main=english]{babel}
\usepackage[dvipsnames]{xcolor}
\usepackage[per-mode=symbol,separate-uncertainty]{siunitx}
\usepackage{graphicx}
\usepackage{dcolumn}
\usepackage{color}
\usepackage{graphicx,wrapfig,lipsum}
\usepackage{bm}
\usepackage{physics}
\usepackage{color}
\usepackage{ulem}
\usepackage{float}
\usepackage{chemformula}

\newcommand{\Ns}{N_\mathrm{S}}
\newcommand{\Nb}{N_\mathrm{B}}
\newcommand{\Cq}{C_\mathrm{q}}
\newcommand{\ar}{\hat{a}_\mathrm{R}}
\newcommand{\ai}{\hat{a}_\mathrm{I}}
\newcommand{\as}{\hat{a}_\mathrm{S}}
\newcommand{\Perr}{P_\mathrm{e}}
\newcommand{\Rq}{R_\mathrm{Q}}
\newcommand{\Rc}{R_\mathrm{C}}

\begin{document}
	\title{Imperfect photon detection in quantum illumination
	}
\author{F.\,Kronowetter}
\email[]{fabian.kronowetter@wmi.badw.de}
\affiliation{Walther-Mei{\ss}ner-Institut, Bayerische Akademie der Wissenschaften, 85748 Garching, Germany} 
\affiliation{Technical University of Munich, TUM School of Natural Sciences, Physics Department, 85748 Garching, Germany}
\affiliation{Rohde \& Schwarz GmbH \& Co. KG, Mühldorfstraße 15, 81671 Munich, Germany}

\author{M.\,Würth}
\altaffiliation[F.\,Kronowetter and M.\,Würth contributed equally to this work]{}
\affiliation{Technical University of Munich, TUM School of Computation, Information and Technology, 80290 Munich, Germany}

\author{W.\,Utschick}
\affiliation{Technical University of Munich, TUM School of Computation, Information and Technology, 80290 Munich, Germany}

\author{R.\,Gross}
\affiliation{Walther-Mei{\ss}ner-Institut, Bayerische Akademie der Wissenschaften, 85748 Garching, Germany}
\affiliation{Technical University of Munich, TUM School of Natural Sciences, Physics Department, 85748 Garching, Germany}
\affiliation{Munich Center for Quantum Science and Technology (MCQST), 80799 Munich, Germany}

\author{K.\,G.\,Fedorov}
\email[]{kirill.fedorov@wmi.badw.de}
\affiliation{Walther-Mei{\ss}ner-Institut, Bayerische Akademie der Wissenschaften, 85748 Garching, Germany}
\affiliation{Technical University of Munich, TUM School of Natural Sciences, Physics Department, 85748 Garching, Germany}
\affiliation{Munich Center for Quantum Science and Technology (MCQST), 80799 Munich, Germany}

\date{\today}
\begin{abstract}
In quantum illumination, various detection schemes have been proposed for harnessing remaining quantum correlations of the entanglement-based resource state. To this date, the only successful implementation in the microwave domain~\cite{Assouly.2023} relies on a specific mixing operation of the respective return and idler modes, followed by single-photon counting in one of the two mixer outputs. We investigate the performance of this scheme for realistic detection parameters in terms of detection efficiency, dark count probability, and photon number resolution. Furthermore, we take into account the second mixer output and investigate the advantage of correlated photon counting (CPC) for a varying thermal background and optimum post-processing weighting in CPC. We find that the requirements for photon number resolution in the two mixer outputs are highly asymmetric due to different associated photon number expectation values.
\end{abstract}

\maketitle

\section{Introduction} 
Nonclassical correlations in propagating signals provide the essential ingredient for quantum illumination (QI)~\cite{Lloyd.2008, Tan.2008}. In QI, a general application scenario is the presence detection of a low-reflectivity object embedded in a bright thermal background. For that purpose, one mode of the entangled resource state is sent as a probe signal, while the other mode is preserved for further use in a joint detection step~\cite{Tan.2008}. Most importantly, QI is robust against an entanglement-breaking background noise, which results in an enhanced performance compared to the ideal classical reference scheme based on coherent-state transmission. Propagating two-mode squeezed vacuum states represent an ideal resource for QI and can be routinely generated with various superconducting parametric circuits \cite{Kraus.2004, Menzel.2012, Pogorzalek.2019, Fedorov.2021}. An optimal detector layout remains an open question, because the full 6 dB quantum advantage (QA) in the error exponent requires very cumbersome and demanding experimental setups~\cite{Zhuang.2017, Shi.2022}. However, a 3 dB quantum advantage over the ideal classical radar can be achieved by using more practically accessible schemes. Among those are the parametric mixer (PM) and phase-conjugate receivers, which both rely on single-photon detection or counting as a final step~\cite{Guha.2009}. Until today, the only experimental implementation of a microwave quantum radar achieving a genuine quantum advantage relies on the PM-type receiver~\cite{Assouly.2023}.

\begin{figure*}
	        \centering
    	    \includegraphics{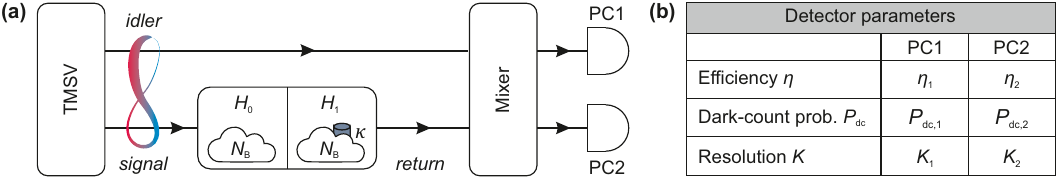}
	        \caption{(a) Illustration of the considered quantum illumination scheme. One mode of a two-mode squeezed vacuum state (TMSV), characterized by $N_\mathrm{S}$ photons per entangled mode, serves as a signal source. While the idler mode can directly pass to the PM receiver, the signal propagates through a bright thermal background characterized by the number of noise photons $N_\mathrm{B}$. Under hypothesis $H_0$ (no target present), the entire signal is lost and only $N_\mathrm{B}$ thermal photons enter the receiver via the signal path. Under $H_1$ (target present), the signal is weakly reflected from the target (with reflectivity $\kappa \ll 1$) and $N_\mathrm{B}+\kappa N_\mathrm{S}$ photons enter the receiver via the signal path. The receiver consists of a mixer, implementing an interaction between the retained idler and return signal modes, followed by two single-photon counters, PC1 and PC2. (b) The PCs can be characterized by their detection efficiency $\eta$, dark count probability $P_\mathrm{dc}$, and photon number resolution $K$. We analyze the performance of the overall scheme in terms of error probabilities for realistic values of $\eta$, $P_\mathrm{dc}$, and $K$ in order to test robustness of the QI quantum advantage in presently accessible experimental settings.}
\label{fig:Fig_1}
\end{figure*}

In this work, we focus on the PM scheme~[cf. Fig.\,\ref{fig:Fig_1}(a)] in the asymptotic regime with realistic microwave photon detector properties. We start with an ideal scenario of perfect photon counters~(PCs). Next, we analyze the PM receiver performance with non-unity detection efficiencies, non-zero dark count probabilities, and finite photon number resolution~[cf. Fig.\,\ref{fig:Fig_1}(b)]. Furthermore, we consider a scheme with two independent PCs and compare its performance with a post-processing protocol, which takes into account the correlated photon counting~(CPC) results of both PCs~\cite{LasHeras.2017}. We find that the respective expectation values of the two photon number operators are highly asymmetric. This is due to the fact that the photon number expectation value at PC1 is governed by the number of signal photons per mode, $N_\mathrm{S}\ll 1$, while the photon number at PC2 is dictated by the number of thermal background photons, $N_\mathrm{B}\gg 1$. As a consequence, PC1 does not require a photon number resolution, $K > 1$, even for a thermal background of 1000 photons. Conversely, PC2 does not outperform the ideal classical radar in individual detection, even with infinite photon number resolution, $K\rightarrow \infty$. Furthermore, we find that the receiver based on correlated photon counting, CPC, performs better than PC1 alone. This performance enhancement is due to a highly correlated nature of detector (PC1 and PC2) outputs. Independent photon counting shows a similar sensitivity towards finite detection efficiencies as the CPC approach. We investigate a weighting of the individual PC measurement results as a function of the system parameters for CPC and identify an optimum balancing between the PCs. 
\section{Quantum illumination protocol}
The QI scheme relies on quantum-enhanced remote sensing by exploiting quantum correlations between $M$ spatially separated pairs of signal and idler modes, described by the bosonic operators $\hat{a}_\mathrm{S}$ and $\hat{a}_\mathrm{I}$, respectively. These quantum correlations are encoded in pure entangled zero-mean Gaussian states which are fully characterized by the corresponding covariance  matrix
\begin{gather}
\textbf{V}_\mathrm{SI}=\langle [ \hat{a}_\mathrm{S}\,\hat{a}_\mathrm{I}\,\hat{a}^\dagger_\mathrm{S}\,\hat{a}^\dagger_\mathrm{I}  ]^\mathrm{T}[ \hat{a}_\mathrm{S}^\dagger\,\hat{a}_\mathrm{I}^\dagger\,\hat{a}_\mathrm{S}\,\hat{a}_\mathrm{I} ] \rangle\\
    =\begin{bmatrix} (N_\mathrm{S} + 1)\,\mathbb{I}_2 & C_\mathrm{q}\,\boldsymbol{\sigma}_\mathrm{X} \\ C_\mathrm{q}\,\boldsymbol{\sigma}_\mathrm{X} & N_\mathrm{S}\,\mathbb{I}_2 \end{bmatrix},
\end{gather}
where $N_\mathrm{S}$ is the mean photon number of the respective signal and idler modes, $\mathbb{I}_2$ is the two-dimensional identity matrix, the quantity $C_\mathrm{q}=\sqrt{N_\mathrm{S}(N_\mathrm{S}+1)}$ encodes the strength of quantum correlations, and $\boldsymbol{\sigma}_\mathrm{X}$ is the Pauli-X matrix. In this framework, the signal mode interrogates a region of interest, while the idler mode is retained and stored for a round-trip time of the signal. For the task of a binary decision between hypothesis~$H_0$ (target absent) and hypothesis~$H_1$ (target present), the return modes entering the receiving unit are given by
\begin{align}
    {\hat{a}_\mathrm{R} \vert}_{H_0}&=\hat{a}_\mathrm{B},\\
    {\hat{a}_\mathrm{R} \vert}_{H_1}&=e^{i\theta}\sqrt{\kappa}\, \hat{a}_\mathrm{S}+\sqrt{1-\kappa}\,\hat{a}_\mathrm{B},
\end{align}
where $\hat{a}_\mathrm{B}$ represents a thermal state with mean photon number $N_\mathrm{B}=\langle \hat{a}^\dagger_\mathrm{B} \hat{a}_\mathrm{B} \rangle \gg 1$, $\theta$ is an overall phase shift and $\kappa \ll 1$ is the round-trip signal loss. Note that, under $H_1$, $N_\mathrm{B}/(1-\kappa)$ thermal photons are encoded in $\hat{a}_\mathrm{B}$ for equal background photon numbers under both hypotheses~\cite{Guha.2009}. Since we focus on QI and do not consider quantum phase estimation, we set $\theta=0$~\cite{Shi.2022}. The resulting joint return-idler state is again characterized by zero-mean Gaussian states with 
\begin{gather}
    \textbf{V}_\mathrm{RI}=\langle [ \hat{a}_\mathrm{R}\,\hat{a}_\mathrm{I}\,\hat{a}^\dagger_\mathrm{R}\,\hat{a}^\dagger_\mathrm{I}  ]^\mathrm{T}[ \hat{a}_\mathrm{R}^\dagger\,\hat{a}_\mathrm{I}^\dagger\,\hat{a}_\mathrm{R}\,\hat{a}_\mathrm{I} ] \rangle\\
    \overset{H_0}{=} \begin{bmatrix} N_\mathrm{B} + 1  & 0 & 0 & 0 \\  0 & N_\mathrm{S} + 1 & 0 & 0  \\  0 & 0 & N_\mathrm{B} & 0 \\  0 & 0 & 0 & N_\mathrm{S} \end{bmatrix} \label{eq:VH0}\\
    \overset{H_1}{=} \begin{bmatrix} \kappa N_\mathrm{S} +N_\mathrm{B} + 1  & 0 & 0 & \sqrt{\kappa}C_\mathrm{q} \\  0 & N_\mathrm{S} + 1 & \sqrt{\kappa}C_\mathrm{q} & 0  \\  0 & \sqrt{\kappa}C_\mathrm{q} & \kappa N_\mathrm{S} +N_\mathrm{B} & 0 \\  \sqrt{\kappa}C_\mathrm{q} & 0 & 0 & N_\mathrm{S} \end{bmatrix} \label{eq:VH1}.
\end{gather}
The minimum error probability, $P_{\mathrm{e,min}}$, for this binary decision task is upper-bounded by the quantum Chernoff bound (QCB), $P_{\mathrm{e,min}}\leq 0.5 \exp(-\Rq M)$, which is asymptotically tight for $M \rightarrow \infty$ and connected to the error exponent $\Rq$. A typical classical reference scheme is a coherent state (CS) transmitter with a mean photon number $\Ns$ per mode, which achieves an error exponent $\Rc=\kappa \Ns/(4\Nb)$. In the weak transmission ($\Ns \ll 1$), bright background ($\Nb \gg 1$), and high loss ($\kappa \ll 1$) limit, the QI protocol has an error exponent $\Rq=\kappa \Ns/\Nb$ which is \SI{6}{dB} larger than $\Rc$ \cite{Tan.2008}.

The goal of various proposed receiver schemes is to reach this theoretical \SI{6}{dB} QA. From Eq.\,(\ref{eq:VH0}) and Eq.\,(\ref{eq:VH1}), it is obvious that the potential of the entanglement-assisted protocol does not stem from local properties of the individual modes (i.e., the diagonal $\textbf{V}_\mathrm{RI}$ entries), but rather from remaining non-local correlations (i.e., the anti-diagonal $\textbf{V}_\mathrm{RI}$ entries) characterized by $\langle \ar \as \rangle=\langle \ar^\dagger \as^\dagger \rangle=\sqrt{\kappa}\Cq$ with $\left[\ar,\as\right]=\left[\ar^\dagger,\as^\dagger\right]=0$. This fundamental result leads to the intuition that a joint measurement of $\ar$ and $\ai$ is a prerequisite for achieving the QA~\cite{Guha.2009, LasHeras.2017, Tan.2008}. A potential workaround for this joint measurement might be implemented with a feed-forward heterodyne scheme, which also avoids using single-photon detectors or counters \cite{Reichert.2023}. Moreover, a variant of this scheme exploiting re-programmable beam splitters promises the full \SI{6}{dB} QA~\cite{Shi.2022}. However, its experimental implementation remains to be very challenging in the microwave regime due to the absence of required components.

\section{Results and Discussion} 
\subsection{Ideal PM receiver characteristics}\label{sec:ideal-OPAR}
 As of today, the only successful experimental implementation of a microwave quantum radar relies on the PM-type receiver, schematically shown in Fig.\,\ref{fig:Fig_1}(a) \cite{Assouly.2023}. In this approach, the return and idler modes interact nonlinearly forming the input-output relations
 \begin{align}
     \hat{b}_1 &= \sqrt{G}\,\ai + \sqrt{G-1}\,\ar^\dagger,\\
     \hat{b}_2 &= \sqrt{G}\,\ar + \sqrt{G-1}\,\ai^\dagger,
 \end{align}
where $G=1+\varepsilon^2$ is the mixer gain and $\varepsilon \ll 1$. Originally, it was proposed to use an optical parametric amplifier for implementing this input-output relation~\cite{Guha.2009}. In the microwave domain, a Josephson ring modulator or a degenerate Josephson mixer realize the same transformation~\cite{Assouly.2023, Renger.2021, Kronowetter.2023}. The optimal mixer gain, $G^*$, has been derived in Ref.~\cite{Shi.2022} as a function of the system parameters, $\Ns$, $\Nb$, and $\kappa$. The mixer is followed by single-photon counters, PC1 and PC2, with
\begin{multline} \label{eq:N1}
    N_1=\langle \hat{b}^\dagger_1 \hat{b}_1 \rangle = G \langle \ai^\dagger \ai \rangle+\\ + \sqrt{G\left(G-1\right)}\left( \langle \ar^\dagger \ai^\dagger \rangle + \langle \ar\ai \rangle  \right)    + \left( G-1 \right) \langle \ar\ar^\dagger \rangle,
\end{multline}
and
\begin{multline}\label{eq:N2}
    N_2=\langle \hat{b}^\dagger_2 \hat{b}_2 \rangle = G \langle \ar^\dagger \ar \rangle+\\ + \sqrt{G\left(G-1\right)}\left( \langle \ar^\dagger \ai^\dagger \rangle + \langle \ar \ai \rangle  \right)
    + \left( G-1 \right) \langle \ai \ai^\dagger \rangle,
\end{multline}
\begin{figure}
    \centering
    \includegraphics{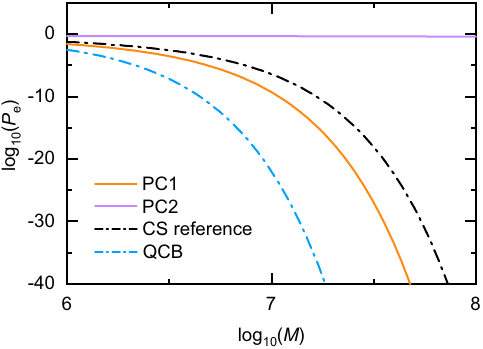}
    \caption{Error probability $\Perr$ as a function of the number of transmitted modes $M$ for individual single-photon counting with PC1 (orange) and PC2 (purple) in an PM-type QI scheme. The dot-dashed black line represents the error probability of the ideal classical reference radar (CS transmitter with a coherent photon number $\Ns$ in combination with a homodyne detector), which coincides with the Helstrom (lower) bound for classical state transmitters \cite{Sorelli.2022}. The dot-dashed blue line depicts the quantum Chernoff (upper) bound for entanglement-based QI schemes. The employed system parameters are $\Ns=0.01$, $\Nb=20$, $\kappa=0.01$, and $G=G^*$.}
    \label{fig:Fig_2}
\end{figure}
as the respective photon numbers. As it can be seen from Eq.\,(\ref{eq:VH0}) and Eq.\,(\ref{eq:VH1}), the return mode $\ar$ is characterized by $\langle \ar^\dagger \ar \rangle=\Nb$ under $H_0$ and $\langle \ar^\dagger \ar \rangle=\Nb + \kappa \Ns$ under $H_1$. The non-local correlations, $\langle \ar^\dagger \ai^\dagger \rangle=\langle \ar \ai \rangle=0$ vanish for $H_0$ and are given by $\langle \ar^\dagger \ai^\dagger \rangle=\langle \ar \ai \rangle=\sqrt{\kappa}\Cq$ for $H_1$. 

All detection protocols perform a maximum likelihood analysis of the photon number statistics, which is based on photon counting through $M$ return-idler transmitted modes. Irrespective of the detection scheme (individual or CPC), it is assumed that for both hypotheses the conditional distributions converge to a normal distribution for $M\gg1$ according to the central limit theorem. The decision threshold for the maximum likelihood test is given by~\cite{Sorelli.2022}
\begin{equation}
    N_\mathrm{th}=M \frac{\sigma|_{H_1}\,\mu|_{H_0} + \sigma|_{H_0}\,\mu|_{H_1}} {\sigma|_{H_0} + \sigma|_{H_1}},\label{eq:ML-threshold}
\end{equation}
where $\mu|_{H_{0,1}}$ and $\sigma^2|_{H_{0,1}}$ are the mean and variance of the photon number distribution under the two hypotheses, respectively. The obtained overlap of the two distributions defines the error probability of the scheme. This overlap depends on the difference of the means $\Delta \mu_i=|N_i\vert _{H_1} - N_i\vert _{H_0}|$ ($i=1,2$), as well as on the respective variances $\sigma_i^2\vert_{H_{0,1}}=N_i(N_i+1)\vert_{H_{0,1}}$, where both quantities scale linearly with the total number of transmitted modes~\cite{Guha.2009}. Since $ N_1\vert_{H_{0,1}} \ll 1 $, it follows that $\sigma_1^2\vert_{H_{0,1}}\approx N_1 \ll 1 $, which results in a small overlap of the two distributions and a low resulting error probability (cf.\,Fig.\,\ref{fig:Fig_2}). Conversely, for the same parameters, $ N_2 \approx 20.104\vert _{H_0}$ ($ N_2  \approx 20.105\vert _{H_1}$) is dominated by $G\langle \ar^\dagger \ar \rangle\approx \Nb$ and yields photon number distributions with variances $\sigma_2^2\vert_{H_{0,1}}\approx N_2^2 \gg 1 \gg \sigma_1^2\vert_{H_{0,1}}$, while $\Delta N_2 \approx \Delta N_1$. The associated error probability of $N_2$ is much larger than that of $N_1$ and clearly inferior to the ideal classical reference scheme, as shown in Fig.\,\ref{fig:Fig_2}. To conclude, the PM-type receiver shows a strong asymmetry of the detection performance in individual detection, where only PC1 (in our convention) shows a QA.

\begin{figure}
    \centering
    \includegraphics{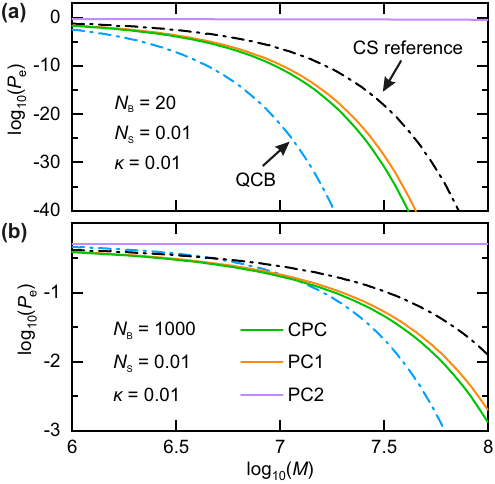}
    \caption{Error probability $\Perr$ as a function of transmitted modes $M$ for individual single-photon detection and correlated photon counting with $w_1=G$ and $w_2=G-1$ for (a) $N_\mathrm{B}=20$ coupled background photons and (b)  $N_\mathrm{B}=1000$. The black dash-dotted line represents the error probability of the ideal CS radar. The blue dash-dotted line depicts the corresponding QCB. The green line depicts the ideal CPC performance, the orange (purple) line shows the results for individual detection with PC1 (PC2). The employed system parameters are $\Ns=0.01$, $\kappa=0.01$, and $G=G^*$.}
\label{fig:Fig_3}
\end{figure}

\subsection{(Un)balanced difference detection}
Las Heras\,\textit{et al.}\,\cite{LasHeras.2017} consider the PM-type receiver exploiting both PCs by analyzing the operator $\hat{O} = G \hat{N}_1  - \left( G-1 \right) \hat{N}_2$. Note that this difference detector (CPC in our convention) is unbalanced with weights $G$ and $G-1$, such that the decisive non-local correlations persist in $\hat{O}$ [cf.~Eq.~(\ref{eq:N1}) and Eq.~(\ref{eq:N2})]. The PCR scheme also utilizes the measurement outcome of both single-photon counters in a balanced difference detector with $\hat{N} = \hat{N}_1-\hat{N}_2$~\cite{Guha.2009}. In the following, we compare the performance of the PM-type scheme in individual photon counting with the CPC approach which implements the operator $\hat{O}$ (see\,Fig.\,\ref{fig:Fig_3}). The benefit of the CPC with respect to individual detection (PC1) is similar for $N_\mathrm{B}=20$ [Fig.\,\ref{fig:Fig_3}(a)] and $N_\mathrm{B}=1000$ [Fig.\,\ref{fig:Fig_3}(b)]. Large correlations between individual photon counting events of PC1 and PC2, illustrated by Eqs.\,(\ref{eq:covH0}) and (\ref{eq:covH1}), result in the enhanced performance of the CPC. The operator $\hat{O}$ can be more generally described as
\begin{equation}
    \hat{O} = w_1 \hat{N}_1 - w_2 \hat{N}_2,
\end{equation}
where $w_1$ and $w_2$ are the weights of the measured photon numbers in post-processing, such that they are not constricted by experimental conditions. The expectation value and variance of $\hat{O}$ are given by
\begin{equation}
    \langle \hat{O} \rangle = w_1\langle \hat{N}_1 \rangle - w_2 \langle \hat{N}_2 \rangle,
\end{equation}
and
\begin{multline}
    \mathrm{Var}\left(\hat{O} \right) = w_1^2\,\mathrm{Var}\left(\hat{N}_1 \right) + w_2^2\, \mathrm{Var}\left(\hat{N}_2 \right) \\
    - 2 w_1 w_2\mathrm{Cov}\left(\hat{N}_1 , \hat{N}_2 \right),
\end{multline}
where $\mathrm{Var}(\hat{N}_i ) = N_i\left(N_i + 1 \right)$ for $i=1,2$.
\begin{figure}
	        \centering
    	    \includegraphics{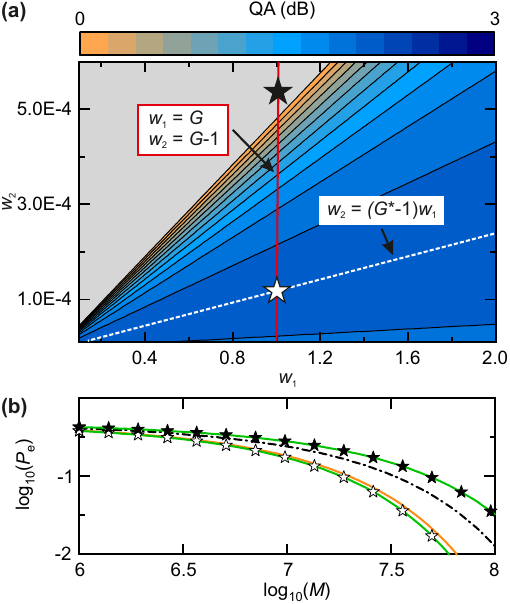}
	        \caption{(a) QA as a function of weights $w_1$ and $w_2$. The solid red line represents weighting according to $w_1=G$ and $w_2=G-1$ \cite{LasHeras.2017}. While this relation (white star) can yield the full \SI{3}{dB} QA (blue color code), already a slight variation or misestimation of $G$ on the order of $0.05\%$ (black star) leads to inferior results with respect to the CS reference. The dashed white line represents an optimal weighting according to Eq.~\ref{eq:opt}. (b) Resulting error probability as a function of transmitted modes for the optimal weighting (green line with white stars) versus a misestimated weighting (green line with black stars) in comparison with individual detection (PC1, orange) and the CS reference (dash-dotted black). The system parameters are $\Ns=0.01, \kappa=0.01, \Nb=1000$, and $G=G^*$.}
\label{fig:Fig_4}
\end{figure}
Under the hypotheses $H_0$ and $H_1$, the corresponding covariances yield
\begin{equation}\label{eq:covH0}
    \left.\mathrm{Cov}\left(\hat{N}_1 , \hat{N}_2 \right)\right\vert_{H_0}=    G\left( G-1 \right)\,\left( N_\mathrm{S} + N_\mathrm{B} + 1 \right)^2,
\end{equation}
\begin{multline}\label{eq:covH1}
    \left.\mathrm{Cov}\left(\hat{N}_1 , \hat{N}_2 \right)\right\vert_{H_1}=    \Bigl( \left( 2G-1 \right)C_\mathrm{q}+\\
    + \sqrt{G\left( G-1 \right)}\,\left( N_\mathrm{S}\left( \kappa + 1 \right) + N_\mathrm{B} + 1 \right)\Bigr)^2.
\end{multline}
Figure~\ref{fig:Fig_4}(a) shows the QA of the CPC approach as a function of $w_1$ and $w_2$. Here, the conventional weighting according to Ref.~\cite{LasHeras.2017}, $w_1=G$ and $w_2=G-1$ (solid red line), has a large gradient with respect to the optimal working regime (blue), such that a small change or misestimation of $G$ can lead to a complete loss of the QA [white star versus black star, see also Fig.\,\ref{fig:Fig_4}(b)]. We identify an optimal weighting according to
\begin{equation}\label{eq:opt}
    \begin{split}
    w_2&=\frac{\sqrt{\Ns(\Ns + 1)(\Nb+\kappa \Ns)(\Nb + \kappa \Ns + 1)}}{(\Nb+(\kappa-1)\Ns)(\Nb + (\kappa+1)\Ns + 1)}\\
     &+\frac{\Ns(\Ns + 1)}{(\Nb+(\kappa-1)\Ns)(\Nb + (\kappa+1)\Ns + 1)}\,w_1\\
     &= (G^*-1)\,w_1,   
    \end{split}
\end{equation}
shown as the dashed white line in Fig.~\ref{fig:Fig_4}(a), which relaxes the requirements in terms of parameter precision and introduces a degree of freedom in the choice of the post-processing weights.

\subsection{Practical PM receiver characteristics}
Realistic quantum illumination implementations may contain detection imperfections~\cite{Assouly.2023, Zhang.2015}. Here, we analyze practical limitations of microwave PCs and their impact on the PM-type receiver. Although single-photon detection for propagating microwaves is challenging due to the low photon energies, which are approximately 5 orders of magnitude smaller than for optical photons, various theoretical concepts~\cite{Romero.2009, Helmer.2009, Koshino.2013, Sathyamoorthy.2014, Fan.2014, Kyriienko.2016, SankarRamanSathyamoorthy.2016, XiuGu.2017, Wong.2017, Leppakangas.2018, Royer.2018} have paved the road to successful experimental implementations~\cite{Chen.2011, Inomata.2016, Besse.2018, Kono.2018, Narla.2016, Lescanne.2020, Dassonneville.2020}. Here, most advanced schemes exploit Ramsey interferometry to implement quantum non-demolition detection, or counting, of incident microwave photons by measuring a photon-induced phase perturbation of an ancilla qubit~\cite{Besse.2018, Kono.2018}. Since the qubit coherence time directly correlates with the dark count rate, the performance of Ramsey-based detectors strongly depends on a sufficiently long qubit lifetime. Apart from the dark count rate and detection efficiency, the photon-number resolution is another key parameter in single-photon detection. Dassonneville\,\textit{et al.}\,\cite{Dassonneville.2020} have realized a number-resolving photon counter of up to three photons, which we consider in our analysis as a reference~(cf.\,Sec.\,\ref{sec:finRes}). 

\subsubsection{Finite detection efficiencies}
\begin{figure}
	        \centering
    	    \includegraphics{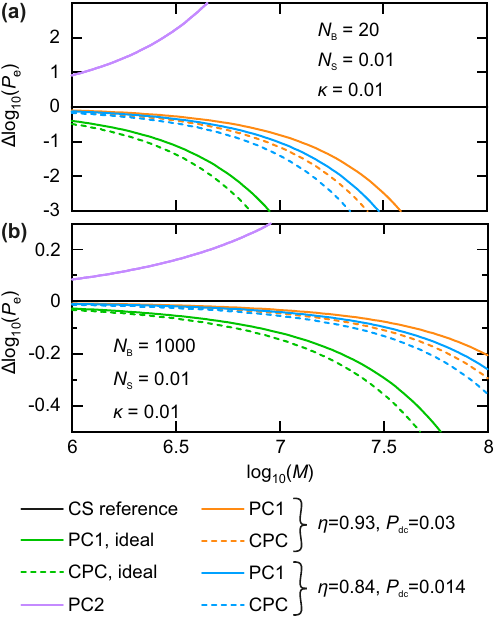}
	        \caption{Error probability difference $\Delta \mathrm{log}_{10}(\Perr)=\mathrm{log}_{10}(\Perr)-\mathrm{log}_{10}(P_\mathrm{e,CS})$ as a function of the number of transmitted modes $M$ for individual photon counting (solid lines) and optimal (according to Eq.\,\ref{eq:opt}) correlated photon counting (dashed lines) for (a) $N_\mathrm{B}=20$ coupled background photons and (b)  $N_\mathrm{B}=1000$. The black line represents the error probability of the CS reference. Green lines depict ideal detectors, orange lines correspond to $\eta=0.93$ and $P_\mathrm{dc}=0.03$, blue lines show the performance for $\eta=0.84$ and $P_\mathrm{dc}=0.014$. The purple line shows the results for PC2 which are close to $P_\mathrm{e}=0.5$ in each of the three considered cases. The mixer gain is set to $G=G^*$.}
\label{fig:Fig_5}
\end{figure}
State-of-the art microwave single-photon detectors (SPDs) achieve a click probability $P_\mathrm{c} = 0.93$ for an incoming single photon with a dark count probability $P_\mathrm{dc} = 0.03$~\cite{Dassonneville.2020}. The dark count probability can be computed as the dark count rate times the duration of the detection window. Moreover, $P_\mathrm{c}=1-P_{\ket{1}}(0)$, where $P_{\ket{1}}(0)$ is the probability of measuring no click for an impinging single photon. To this date, the quality of photon-number resolved measurements, expressed by the conditional probabilities $P_{\ket{k}}(l)$ of realizing a measurement outcome $l=\{0,\ldots,3\}$ for an incoming Fock state $\ket{k}$ strongly depends on $k$, with $P_{\ket{1}}(1)= 76 \%, P_{\ket{2}}(2)= 71 \%$ and $P_{\ket{3}}(3)= 54 \%$~\cite{Dassonneville.2020}. For simplicity, we assume a photon number resolution corresponding to the SPD click probability, i.e., $P_{\ket{k}}(k)=P_\mathrm{c}=0.93$ for all $k=1,2,...$ which effectively gives an upper performance bound. We model the influence of finite dark count probabilities and a finite detection efficiency $P_{\ket{k}}(k)<1$ with a beam splitter before an ideal PC
\begin{equation}
    \hat{c}_i=\sqrt{\eta}\, \hat{b}_i+ \sqrt{1-\eta}\, \hat{b}_{\mathrm{c},i},
\end{equation}
where $i=1,2$, $\eta=P_{\ket{k}}(k)$ is the beam splitter transmissivity, the dark count probability is modeled with a coupled mode $\hat{b}_{\mathrm{c},i}=1/\sqrt{1-\eta} \, \hat{a}_{\mathrm{th},i}$ characterized by $\langle \hat{a}^\dagger_{\mathrm{th},i} \hat{a}_{\mathrm{th},i} \rangle=\left(1/(1-P_\mathrm{dc})-1 \right)\approx P_\mathrm{dc}$ for $P_\mathrm{dc}\ll 1$. Accordingly, we do not take into account the influence of $P_{\ket{k}}(l)$ for $k\neq l$.
\begin{figure*}
	        \centering
    	    \includegraphics{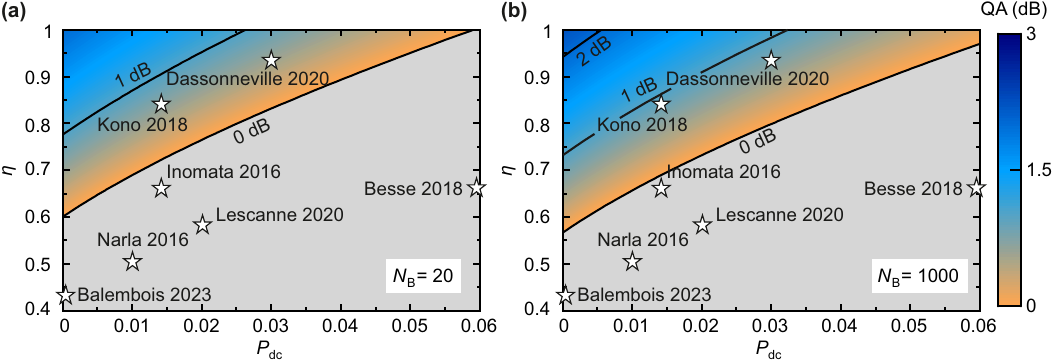}
	        \caption{QA for individual detection with PC1 as a function of the dark count probability $P_\mathrm{dc}$ and detection efficiency $\eta$ for (a) $\Nb=20$ and (b) $\Nb=1000$. The white stars indicate the performances of microwave single-photon detector implementations from Refs.\,\cite{Kono.2018, Dassonneville.2020, Inomata.2016, Besse.2018, Lescanne.2020, Narla.2016, Balembois.2023}. The gray color code illustrates a region without any QA. The other employed system parameters are $\Ns=0.01$, $\kappa=0.01$ and $G=G^*$.}
\label{fig:Fig_6}
\end{figure*}

In Fig.\,\ref{fig:Fig_5}, we plot the error probability difference for $N_\mathrm{B}=20$ [Fig.\,\ref{fig:Fig_5}(a)] and for $N_\mathrm{B}=1000$ [Fig.\,\ref{fig:Fig_5}(b)]. We compare the ideal performance of individual and CPC detection for a given CS reference in two realistic scenarios: high efficiency and moderate dark count probability~\cite{Dassonneville.2020} versus moderate efficiency and low dark count probability~\cite{Kono.2018}. For $N_\mathrm{B}=20$, both scenarios yield clearly inferior results compared to the ideal case. We observe that the scenario with low dark count probability outperforms the high-efficiency counterpart, which underlines that for a realistic implementation, minimizing dark count probabilities plays a decisive role for the photon detection in QI. Additionally, the CPC approach only marginally beats individual detection for all three scenarios in the low-noise case of $N_\mathrm{B}=20$. The high-noise regime, $N_\mathrm{B}=1000$, shows similar results with the low dark count detector performing slightly better than the high efficiency case and a consistent performance enhancement for the CPC. In both noise regimes, PC2 does not exhibit a strong dependence on the non-idealities, since $\Perr \approx 0.5 $ already in the ideal case.

Figure\,\ref{fig:Fig_6} illustrates the performance of various already demonstrated microwave single-photon detectors, in terms of $P_\mathrm{dc}$ and  $\eta$, for achieving a QA in individual detection with PC1. In accordance with the successful microwave quantum radar realization~\cite{Assouly.2023}, the underlying device investigated by Dassonneville \textit{et al.}~\cite{Dassonneville.2020} is situated in the region of a robust QA. Importantly, the QA vanishes rapidly with increasing $P_\mathrm{dc}$, even for an ideal detection efficiency of $\eta=1$. Conversely, the scheme is robust against finite efficiencies, $\eta<1$, down to $\eta \approx 0.6$ for $N_\mathrm{B}\gg 1$ [cf.~Fig.\,\ref{fig:Fig_6}]. These findings suggest that the minimization of $P_\mathrm{dc}$ in combination with a reasonably high $\eta$ are desirable in order to achieve the QA, in agreement with our results from Fig.\,\ref{fig:Fig_5}. In accordance with theory, the maximally reachable QA increases with increasing $N_\mathrm{B}$, as it also can be seen in Fig.\,\ref{fig:Fig_6}(a) and (b). As a consequence, the area of $\mathrm{QA}>0$ increases and, e.g., the \SI{1}{dB} and \SI{0}{dB} QA lines lean towards lower values of $P_\mathrm{dc}$ and $\eta$. We would like to mention that the mapping of existing single-photon detectors onto the QA problem in Fig.\,\ref{fig:Fig_6} may be limited to various simplifications of our theoretical model and should not be considered as an overall evaluation of those detectors' performance.

\subsubsection{Finite detection resolution}\label{sec:finRes}

In principle, photon counters can provide a full access to the photon number operator in Eq.\,(\ref{eq:N1}) and Eq.\,(\ref{eq:N2}). However, existing state-of-the-art microwave single-photon counters exhibit a rather limited photon number resolution~\cite{Kono.2018,Lescanne.2020,Dassonneville.2020}. To analyze the impact of this finite resolution, we restrict the PCs in Fig.\,\ref{fig:Fig_1} to a resolution up to $K$ photons. Therefore, a single measurement has possible outcomes of measuring $0,1,\ldots,K$ photons and we assume that the measurement yields $K$ if the number of photons in the probe is larger or equal $K$. Under both hypotheses the state at the output of the mixer is a thermal state with mean photon number $N_i, i\in\{1,2\}$, given by Eqs.\,(\ref{eq:N1},\ref{eq:N2})~\cite{Guha.2009}. This yields a photon number distribution at the output of the photon counter 
\begin{equation}
	p(n) = \begin{cases}
		(1-q_i)q_i^n,	& 0\leq n<K,\\
		1-(1-q_i)\sum_{k=0}^{K-1}q_i^k,	&n=K,
	\end{cases}
\end{equation}
with $q_i = \frac{N_i}{N_i+1}$. For this distribution, the expectation value is given by
\begin{equation}
	\mu_i^{(K)} = \frac{q_i(1-q_i^K)}{1-q_i}, \label{eq:mean-finite}
\end{equation}
and the respective variance can be written as
\begin{multline}
	\sigma_i^{2,(K)} = \frac{q_i}{1-q_i} \left(1-(2K+1)q_i^K + 2q_i\frac{1-q_i^K}{1-q_i}\right) \\
	        - \left(\frac{q_i(1-q_i^K)}{1-q_i}\right)^2. \label{eq:var-finite}
\end{multline}
Analogously to Section~\ref{sec:ideal-OPAR}, the assumption of large number of transmitted modes, i.e. $M\gg1$, leads to the threshold given in Eq.~(\ref{eq:ML-threshold}) but with mean and standard deviation from Eqs.~(\ref{eq:mean-finite},\ref{eq:var-finite}). 

\begin{figure}
    \centering
    \includegraphics{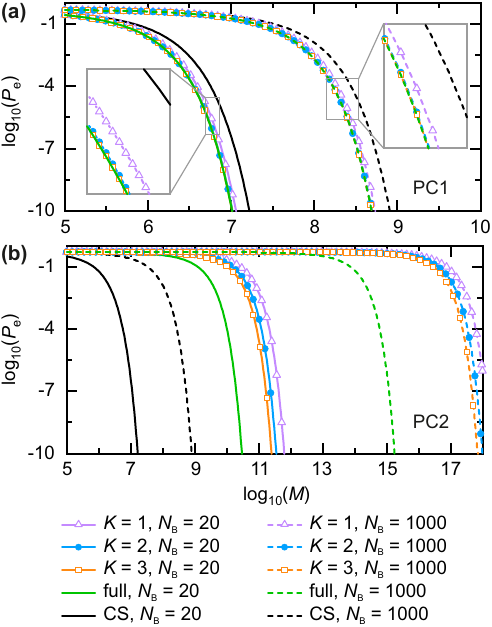}
    \caption{Error probability $\Perr$ as a function of the number of transmitted modes $M$ for different resolutions $K$ for individual detection with (a) PC1 and (b) PC2. The solid lines show the performance for the low-noise regime, $\Nb=20$, and dashed lines for the high-noise regime, $\Nb=1000$. The mean signal photon number is $\Ns=0.01$ and reflectivity $\kappa=0.01$ in both cases. The mixer gain is set to $G=G^*$.}
    \label{fig:Fig_7}
\end{figure}
In Fig.\,\ref{fig:Fig_7}, we plot the error probability for individual detection with PC1 and PC2 in panels (a) and (b), respectively. We compare ideal photon counters with values of $K=1,2,3$. For PC1 and both background scenarios of $\Nb=20$ and $\Nb=1000$, all variants clearly outperform the CS reference and for a resolution of $K\geq2$, the resulting error coincides with the full-resolution counter. The associated low average photon number $N_1 \approx 0.11\ll 1$ under both hypotheses and for both background scenarios explains why a binary SPD is close to being optimal. The reason why a higher resolution does not become more relevant in more noisy scenarios is that the optimal mixer gain, $G^*$, decreases with increasing $N_\mathrm{B}$, such that $N_1$ stays similar for varying $N_\mathrm{B}$. This finding is important for real-world applications with naturally high number of noise photons. Conversely, detection with PC2 alone is always inferior in comparison to the optimum classical scheme due to the large value of $N_2$, which is governed by strong background noise coupled to the return mode~[cf. Eq.~(\ref{eq:N2})].

\section{Conclusion}
Experimental demonstration of a QA in quantum sensing protocols is a demanding task due to stringent requirements on experimental imperfections. As a consequence, successful implementations in the optical regime, as well as at microwave carrier frequencies, so far achieve a QA of only around $\SI{0.8}{dB}$~\cite{Zhang.2015, Assouly.2023} out of potentially obtainable $\SI{3}{dB}$. Here, we have performed a detailed analysis of particular experimental imperfections on the QI performance for the PM-type receiver schemes. In this context, we have focused on the performance of single-photon counters, which represent one of the central elements of QI protocols. We have compared individual single-photon detection to correlated difference detection, and have found that the latter performs slightly better for different noise regimes. We have analyzed the role of respective weighting of the individual detector outcomes and identified the adjusted ratio for the optimal QI performance [cf.~Eq.(\ref{eq:opt})]. While large $N_\mathrm{B}\approx 1000$ theoretically gives access to a larger QA and matches realistic noise values at \si{GHz} frequencies, the detection unit needs to be able to handle corresponding return signal powers. Respectively, the mixer needs to operate at signal powers on the order of $N_\mathrm{B}\cdot \hbar \omega$ without suffering from compression effects; the same applies to PC2 [cf.\,Eq.\,(\ref{eq:N2})]. As a consequence, currently available PCs with a photon number resolution of up to three photons yield a clearly deteriorated performance of PC2 already for small $N_\mathrm{B}\approx 10$, such that CPC does not seem to be suitable for near-term implementations. In contrast, individual detection with PC1 with limited photon number resolution, $K\leq 3$, reaches a reasonable QA even for large $N_\mathrm{B}\approx 1000$. As we have noted, the maximum achievable QA with PM-type receivers increases with increasing $N_\mathrm{B}$ and saturates at \SI{3}{dB} for large $N_\mathrm{B}$, which imposes further restrictions on low-$N_\mathrm{B}$ implementations~\cite{Assouly.2023}. Therefore, individual detection with PC1 for large $N_\mathrm{B}$ may be the simplest route towards the practical \SI{3}{dB} QA with the currently available technology. Finally, our presented results provide valuable insights also to neighboring research fields, such as quantum communication~\cite{Pirandola.2021, Zhang.2023, Fesquet.2022}. While the fundamental hurdle of transmitting quantum signals at microwave frequencies over free-space channels remains to be an ongoing challenge, careful analysis of suitable operation scenarios represents a decisive step towards experimental realization.\\

\begin{acknowledgments}
We acknowledge support by the German Research Foundation via Germany’s Excellence Strategy (EXC-2111-390814868), the German Federal Ministry of Education and Research via the project QUARATE (Grant No.13N15380). This research is part of the Munich Quantum Valley, which is supported by the Bavarian state government with funds from the Hightech Agenda Bayern Plus.
\end{acknowledgments}
	
\bibliography{Bib}

\end{document}